\documentclass[pdflatex,sn-mathphys-num,iicol]{sn-jnl}% Math and Physical Sciences Numbered Reference Style
\usepackage{graphicx}%
\usepackage{multirow}%
\usepackage{amsmath,amssymb,amsfonts}%
\usepackage{amsthm}%
\usepackage{mathrsfs}%
\usepackage[title]{appendix}%
\usepackage{xcolor}%
\usepackage{textcomp}%
\usepackage{manyfoot}%
\usepackage{booktabs}%
\usepackage{algorithm}%
\usepackage{algorithmicx}%
\usepackage{algpseudocode}%
\usepackage{listings}%
\usepackage{bm}%
\usepackage{float}%
\usepackage{subcaption}%
\theoremstyle{thmstyleone}%
\theoremstyle{thmstyletwo}%

\theoremstyle{thmstylethree}%

\begin{document}

\title[Article Title]{Rejection-free Glauber Monte Carlo for the 2D Random Field Ising Model via Hierarchical Probabilistic Counters}

%%=============================================================%%
%% GivenName	-> \fnm{Joergen W.}
%% Particle	-> \spfx{van der} -> surname prefix
%% FamilyName	-> \sur{Ploeg}
%% Suffix	-> \sfx{IV}
%% \author*[1,2]{\fnm{Joergen W.} \spfx{van der} \sur{Ploeg} 
%%  \sfx{IV}}\email{iauthor@gmail.com}
%%=============================================================%%

\author*[1]{\fnm{Luca} \sur{Cattaneo}}\email{luca3.cattaneo@polimi.it}

\author[1]{\fnm{Federico} \sur{Ettori}}\email{federico.ettori@polimi.it}
%\equalcont{These authors contributed equally to this work.}

\author[1]{\fnm{Giovanni} \sur{Cerri}}\email{giovanni1.cerri@mail.polimi.it}
%\equalcont{These authors contributed equally to this work.}

\author[1]{\fnm{Paolo} \sur{Biscari}}\email{paolo.biscari@polimi.it}
%\equalcont{These authors contributed equally to this work.}

\author[1]{\fnm{Ezio} \sur{Puppin}}\email{ezio.puppin@polimi.it}
%\equalcont{These authors contributed equally to this work.}

\affil*[1]{\orgdiv{Department of Physics}, \orgname{Politecnico di Milano}, \orgaddress{\street{Piazza Leonardo da Vinci 32}, \city{Milan}, \postcode{20133}, \country{Italy}}}

%\affil[2]{\orgdiv{Department}, \orgname{Organization}, \orgaddress{\street{Street}, \city{City}, \postcode{10587}, \state{State}, \country{Country}}}

%\affil[3]{\orgdiv{Department}, \orgname{Organization}, \orgaddress{\street{Street}, \city{City}, \postcode{610101}, \state{State}, \country{Country}}}

%%==================================%%
%% Sample for unstructured abstract %%
%%==================================%%

\abstract{We present an efficient Monte Carlo algorithm for the simulation of the two-dimensional Random Field Ising Model (RFIM). The method combines the event-driven, rejection-free character of the Bortz Kalos-Lebowitz (BKL) algorithm with Glauber transition probabilities, introducing hierarchical probabilistic counters to perform spin selection in {\mathversion{normal}$\mathcal{O}(\log N)$} operations. This enables efficient sampling of the system’s dynamics, especially in the low-temperature and low-disorder regime, where traditional Metropolis updates suffer from critical slowing down. Furthermore, this approach allows a proper dynamical simulation of the Ising system's behavior even in the presence of a Random Field (RF), unlike the BKL method. RFIM simulations with Gaussian field distributions reproduce the expected reduction of the pseudo-critical temperature with increasing disorder. Benchmarking shows speedups exceeding two orders of magnitude compared to the Metropolis algorithm in the low-temperature regime. The proposed method provides an efficient and dynamically faithful tool for studying both equilibrium and non-equilibrium phenomena in disordered spin systems.}


%%\pacs[JEL Classification]{D8, H51}

%%\pacs[MSC Classification]{35A01, 65L10, 65L12, 65L20, 65L70}

\maketitle
\section{Introduction}\label{Introduction}

\textit{Monte Carlo} (MC) \textit{methods} are among the most powerful computational techniques for studying statistical systems characterized by a large number of degrees of freedom, such as the \textit{Ising Model} \cite{Ising1925}. Their stochastic nature allows for efficient sampling of high-dimensional configuration spaces, particularly where analytic approaches are intractable. These methods are widely used in statistical mechanics to explore equilibrium and non-equilibrium properties of complex systems.

One of the earliest and most widely adopted Monte Carlo techniques is the \textit{Metropolis algorithm} \cite{Metropolis1953}, which provides a simple yet effective approach to sampling spin configurations according to the Boltzmann distribution. Using a \textit{single spin flip} update, this method ensures the \textit{detailed balance} and \textit{ergodicity} conditions needed in Monte Carlo simulations \cite{Janke2012}. Despite its conceptual simplicity, the Metropolis algorithm can become computationally inefficient near \textit{criticality} (critical slowing down) or at \textit{low temperatures}, where the acceptance rate of proposed moves decreases significantly. 

To address these limitations, the classical \textit{BKL} (or \textit{N-Fold Way}) \textit{algorithm} \cite{Bortz1975} was developed as a rejection-free alternative that accelerates the exploration of configuration space. This event-driven approach not only reduces variance in simulation times but also enhances efficiency in systems where rare events dominate the dynamics. The \textit{N-Fold Way} algorithm is still a cornerstone in computational statistical physics. Its rejection-free nature makes it particularly effective for simulating spin systems characterized by infrequent events, allowing for accurate dynamical studies even in regimes where standard Metropolis dynamics become inefficient. Recent work by Ettori \textit{et al.} \cite{Ettori2024} has demonstrated that the N-Fold Way algorithm continues to play a key role in the computation of nucleation rates and relaxation processes in Ising-like models, particularly in the low-temperature regime where activated dynamics dominate. This regime is of fundamental interest, as it provides insight into metastability, rare-event statistics, and non-equilibrium behavior. However, while the N-Fold Way achieves remarkable accuracy in these systems, its extension to disordered systems such as the \textit{Random Field Ising Model} (RFIM) remains nontrivial, as the algorithm cannot directly account for the complex landscape induced by quenched disorder. Recent developments have introduced algorithmic upgrades to improve both the efficiency and scalability of standard Monte Carlo protocols. For instance, Macias-Medri \textit{et al.} \cite{Macias-Medri2023} and Müller \textit{et al.} \cite{Mller2023} have demonstrated that performance gains can be achieved through algorithmic improvements of the Metropolis method in complex systems. While these approaches provide meaningful speedups in specific settings, they do not alter the underlying rejection-based dynamics of the Metropolis algorithm, which remain the dominant source of inefficiency in the low-temperature regime.

As a direct consequence, the challenge of efficiently simulating Random Field Ising systems at low temperatures remains largely unresolved, because the N-Fold Way algorithm is not able to correctly reproduce the complexity and the effects of the Random Field (RF) on the Ising system.

The two-dimensional Random Field Ising Model is a paradigmatic system for studying disorder in statistical physics \cite{Belanger1991}. 
In the thermodynamic limit, the 2D RFIM does not sustain spontaneous long-range ferromagnetic order for any finite disorder strength $\sigma>0$ \cite{Imry1975,Aizenman1989,Binder1983,Bricmont1987,Aizenman2019}.
Nevertheless, finite-size systems display rich behavior, including the formation of large correlated spin clusters, long-lived metastable states, slow relaxation dynamics, and avalanche-like magnetization changes \cite{Metra2021,Fytas2018,Balog2018,Sinha2013}.
These features make low-temperature RFIM simulations both physically relevant and computationally challenging, motivating the development of efficient algorithms capable of accurately capturing its nontrivial dynamics.

There have been notable advancements in other, non-Monte-Carlo, methods for RFIM simulations, in particular with the work by Fytas and Martin-Mayor \cite{Fytas2016}. However, this method is limited to the zero temperature analysis of RFIM and therefore does not directly apply to the finite-temperature regime considered here.

Therefore, to study the RFIM at low temperature, where the Metropolis algorithm undergoes many spin flip rejections, novel algorithms capable of overcoming the limitations of both traditional rejection-based and rejection-free methods are needed. 

Among the alternative strategies developed to overcome the intrinsic limitations of single-spin-flip dynamics, \textit{Parallel Tempering} (PT) or \textit{Replica Exchange Monte Carlo} \cite{Hukushima1996} has emerged as a powerful technique to improve equilibration in rugged energy landscapes. By simultaneously simulating multiple replicas of the system at different temperatures and allowing periodic exchanges between them, PT effectively alleviates trapping in local minima and accelerates convergence toward equilibrium. However, its efficiency depends critically on the overlap between energy distributions at adjacent temperatures, which becomes increasingly difficult to achieve in disordered systems such as the RFIM, where local RFs generate highly heterogeneous energy barriers that hinder replica exchanges. \textit{Cluster Monte Carlo} algorithms, such as the \textit{Swendsen–Wang} \cite{Swendsen1987} and \textit{Wolff} \cite{Wolff1989} methods, offer another route to mitigate critical slowing down by flipping correlated spin clusters instead of individual spins. These methods have proven extremely effective in pure Ising and Potts models. Nevertheless, their efficiency collapses in the presence of quenched disorder, as the random local fields break the symmetry required to define energetically favorable clusters, and thus the cluster construction becomes ill-defined. \textit{Parallelization} strategies, including GPU-based implementations and domain decomposition approaches \cite{Preis2009}, have also been proposed to accelerate large-scale Ising simulations. Although these techniques provide substantial computational gains through hardware-level optimization, they do not fundamentally resolve the algorithmic bottlenecks associated with slow dynamics and complex metastable states characteristic of the RFIM. In addition, \textit{Population Annealing} (PA) methods \cite{Machta2010} combine a population of system replicas with annealing schedules and weighted‐averaging to sample rugged free‐energy landscapes more effectively. However, although PA is well suited to parallelization and shows promise in glassy systems, its practical application to the RFIM remains challenging: the presence of quenched RFs and the large number of metastable states typical of the RFIM complicate the population‐reweighting step and hinder efficient sampling across the full low-temperature phase space. 

Therefore, despite the complexity and success of these modern Monte Carlo techniques, none of them efficiently captures the low-temperature dynamics in the presence of a RF, reinforcing the need for novel algorithmic frameworks capable of addressing these intrinsic challenges. 

In this work, we present a new Monte Carlo approach combining the event-driven character of the BKL algorithm with probabilistic counters enabling $\mathcal{O}(\log N)$ spin selection under Glauber dynamics. This method preserves dynamical fidelity while yielding substantial performance gains in the low-temperature regime. 

The outline of the work is the following. In Sec. \ref{sec: The BKL or N-Fold Way Algorithm}, we briefly review the BKL algorithm. In Sec. \ref{sec: The algorithm}, we describe in detail the idea behind the present algorithm and its implementation. In Sec. \ref{sec: Validation and Performance Comparison with Metropolis}, we first study the correct functioning of the previously introduced algorithm by applying it both to the pure 2D Ising model and to the 2D RFIM to observe their characteristic behaviors around the critical temperature of the pure Ising case. We then study the gain in computational time given by the present algorithm in a wide temperature range for the 2D RFIM, with different parameters of the RF distribution and different lattice sizes. Finally, we make some concluding comments in Sec. \ref{sec: Conclusions}.

%The Introduction section, of referenced text \cite{bib1} expands on the background of the work (some overlap with the Abstract is acceptable). The introduction should not include subheadings.

%Springer Nature does not impose a strict layout as standard however authors are advised to check the individual requirements for the journal they are planning to submit to as there may be journal-level preferences. When preparing your text please also be aware that some stylistic choices are not supported in full text XML (publication version), including coloured font. These will not be replicated in the typeset article if it is accepted. 

\section{The BKL or N-Fold Way Algorithm}
\label{sec: The BKL or N-Fold Way Algorithm}

While the Metropolis algorithm has proven to be a robust and general-purpose tool for Monte Carlo simulations of systems close to or at equilibrium, it suffers from inefficiencies in systems where the probability of accepting proposed updates is very low. Such situations frequently arise at low temperatures or in systems with large energy barriers between states, where the majority of proposed moves are rejected, resulting in long periods of stasis and poor statistical sampling. Moreover, Metropolis dynamics is typically used for equilibrium sampling and does not provide a well-defined physical time mapping; this limits their usefulness for studies that require a faithful mapping to physical time, as in kinetic Monte Carlo simulations.

To overcome these limitations, Bortz, Kalos, and Lebowitz introduced the so-called \textit{N-fold way algorithm} (also known as the BKL algorithm) in 1975 \cite{Bortz1975}. This method belongs to the class of \emph{rejection-free} or \emph{event-driven} Monte Carlo algorithms and is particularly effective for systems with discrete state spaces, such as spin models.

The central idea of the BKL algorithm is to eliminate the inefficiency caused by rejected moves by selecting, at each time step, an update that is guaranteed to be accepted. This is accomplished by categorizing all possible updates into classes based on their transition rates. For instance, in the context of the Ising model, one can classify spin flip events according to the interaction energy $E_i$ of the selected spin, and compute the transition probability $p_i$ corresponding to the i-th spin $s_i$ using the \textit{Glauber dynamics} \cite{Glauber1963}:
\begin{equation}\label{eq.GlauberP}
p_i = \frac{1}{2\alpha} \left( 1 - \tanh{\left(\beta E_i\right)}\right),
\end{equation}
where $\beta = 1/k_B T$, with $k_B$ fixed at 1 for the sake of simplicity, and $\alpha$ is a microscopic characteristic time for spin flipping. In transition-metal ferromagnets the spin-flip relaxation time is of the order of picoseconds. The parameter $\alpha$ simply rescales the time unit and is therefore set to unity in the following. Using these probability transitions the algorithm is able to properly describe the dynamical evolution of the system.

Then, the algorithm selects a class with probability proportional to $p_i$, and within that class, a specific site is chosen uniformly at random to perform the update.

Time in the BKL algorithm is advanced stochastically, at each spin-flip event, of a quantity:
\begin{equation}\label{eq.Deltat}
\Delta t = -\frac{\tau}{P} \ln r,
\end{equation}
where $\tau = 1$ is the correct choice when transition rates already contains the prefactor $\alpha$, $P = \sum_i p_i$ is the sum over all spins and $r$ is a uniform random number in $(0,1)$. This procedure associates a \textit{physical time} increment to each accepted event and avoids the inefficiency arising from long sequences of rejected updates.

The N-Fold Way achieves a significant speedup in scenarios where accepted events are rare, since each step results in an actual state change. This makes it especially useful in kinetic Monte Carlo simulations of metastable decay, nucleation, and aging phenomena, where long periods of microscopic inactivity can dominate the system's evolution.

Moreover, with the use of the Glauber dynamics, the BKL method maintains exact statistical equivalence with continuous-time Markov processes in the appropriate limit, providing not only computational efficiency but also theoretical rigor. 

Despite its advantages, the BKL algorithm does come with implementation challenges. Maintaining and updating the event classification and rate tables can be computationally demanding, particularly for large or complex systems. Nevertheless, in cases where rejection rates are prohibitively high (i.e., at low temperatures), it remains the most efficient alternative to the traditional Metropolis dynamics.

\section{The algorithm}
\label{sec: The algorithm}

In our simulations, we consider the two-dimensional Random Field Ising Model, defined on a square lattice of size $L\times L$ by the Hamiltonian
\begin{equation}
\mathcal{H} = -J \sum_{\langle i,j\rangle} s_i s_j - \sum_i (H + h_i)s_i ,
\end{equation}
where $s_i=\pm1$ are Ising spins, $J>0$ is the ferromagnetic coupling, $H$ is a uniform external field, and $h_i$ are quenched random fields drawn independently from a Gaussian distribution with zero mean and variance $\sigma^2$, which controls the disorder strength. Periodic boundary conditions are imposed in both lattice directions.

The addition of a RF to the Ising system implies that each spin site has an energy that depends not only on the value of the spin itself and of its neighbors' spin, but also on the on-site realization of the RF. The site-dependent RF value makes it impossible to group spins into a small number of classes, as required by the N-fold way algorithm, since we would need a number of classes equal to the total number of spins in the lattice $N = L \times L$, which removes the advantages of this class-based approach. 

Instead, the Metropolis algorithm, by the way it is conceived, is very general and can also be applied to this RF scenario. However, it tends to be very inefficient at low temperatures, where this algorithm rejects many possible spin flips before actually being able to accept the flip of a spin. Moreover, the Metropolis algorithm does not necessarily implement Glauber transition rates and therefore does not correspond to the physical single-spin-flip dynamics usually assumed for kinetic studies.

For these reasons, it would be much better to borrow the BKL algorithm’s ability to make an \textit{informed choice} of the spin to flip, never rejecting a possible spin flip, with the probability of choosing a spin being proportional to the transition probability $p_i$ of the Glauber dynamics given by eq. \ref{eq.GlauberP}. The algorithmic framework presented here can be straightforwardly formulated using either Glauber or Metropolis transition rates, simply by replacing eq. \ref{eq.GlauberP} with the corresponding acceptance probability. The algorithmic advance itself is therefore not tied to a specific choice of dynamics. We adopt Glauber dynamics for two main reasons. First, Glauber rates provide a well-defined physical time scale, which allows to access both equilibrium properties and nonequilibrium dynamical behavior within a single consistent framework. While this aspect is not particularly relevant for the processes considered in the present work, it will play a central role in future studies of the nonequilibrium behavior and dynamical phase transitions of the RFIM. Second, Glauber dynamics is slightly more computationally demanding than Metropolis dynamics; using it therefore represents a conservative and robust benchmark for evaluating the performance of the proposed algorithm.

The basic idea behind this approach is to use some \textit{hierarchical cumulative counters} to select the spin to flip. Considering a 2-Dimensional RFIM with periodic boundary conditions, we can calculate the interaction energy \textit{$E_i$} of a spin $i$ with its four nearest neighbor spins, the external constant field \textit{H} and the RF \textit{h} using: 
\begin{equation}\label{eq.DeltaE}
    E_i = s_i \left(J\sum_{j=N.N.} s_j + H + h_i\right)
\end{equation}
where the sum is calculated on the 4 Nearest Neighbors spins (NNs) $s_j$ of the $i-th$ spin $s_i$, the Heisenberg exchange interaction factor J between couples of spins is set to 1 for simplicity, and $h_i$ is the realization of the RF at the $i-th$ lattice site (where the spin $i$ is). Then, to select a spin with a probability proportional to its transition probability $p_i$ (given by eq. \ref{eq.GlauberP}), we calculate the sum of all the transition probabilities $p_{tot} = \sum_i p_i $ and assign a region of the $ \left[ 0,p_{tot} \right] $ interval to each spin with an amplitude equal to its transition probability $p_i$. Extracting with uniform probability a random real number belonging to the $ \left[ 0,p_{tot} \right] $ interval, we end up in a segment corresponding to one of the $N$ spins of the system, which is the one selected for the flip, as the next step in the Markov chain. A naive implementation of this idea would consist of building an $N$-long array in which the i-th element contains the sum of the transition probabilities given by eq. \ref{eq.GlauberP}, up to itself included. By doing so, this array contains the values between 0 and $p_{tot}$, the sum of all the transition probabilities of all the spins in the lattice. Then, we could extract a random number in this $\left[0,p_{tot}\right]$ interval. This random value will fall into one of the $N$ intervals that form the array. In this way, the random number selects one of the $N$ spins in the lattice in a way that is proportional to the intervals' amplitude, i.e., the transition probability of each spin. However, to find the spin to flip using a random real number, we should compare it to many of the $N$ elements of this new array (potentially also all of them) and, once flipped, we would have to overwrite all the transition probabilities in the array that comes after the flipped spin. 

\begin{figure}[t]
  \centering
  \includegraphics[width=0.48\textwidth, height=6cm]{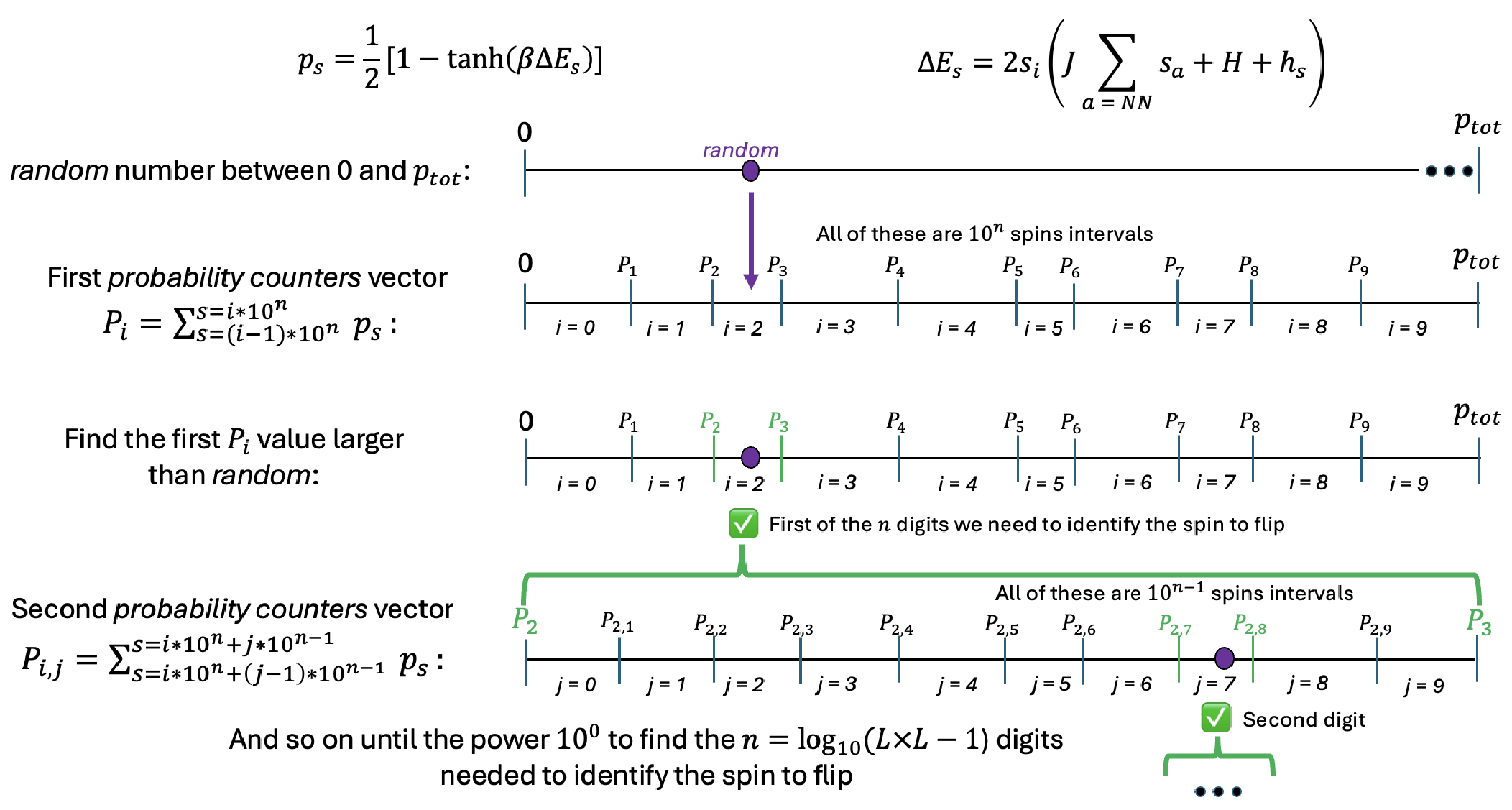}
  \caption{Schematic representation of how the hierarchical cumulative counter vectors are built and used in the present algorithm. Lowercase letters indicate the transition probability of the spins $p_s$, while uppercase letters represent the cumulative counters $P_i$.}
  \label{fig:Flowchart}
\end{figure}

From an algorithmic perspective, this weighted spin-selection problem is related to the class of data structures used to perform prefix-sum queries and weighted random sampling, such as \textit{Fenwick trees} \cite{Fenwick1994}. In this framework, the selection of a spin according to its transition probability can be carried out in $\mathcal{O}(\log N)$ time, while the construction (or full initialization) of the cumulative probability structure requires $\mathcal{O}(N)$ operations. The approach introduced here can be interpreted as a concrete realization of this general idea, adapted to the specific requirements of RFIM dynamics. In particular, the hierarchical cumulative counters provide logarithmic-time access to spin selection while allowing for efficient local updates after each spin flip, since only the transition probabilities of the flipped spin and its nearest neighbors must be modified.

The solution used in our approach to reduce the number of scans needed and to improve efficiency is to set up some \textit{hierarchical cumulative counters} using a base-10 reference. For the lattice sizes considered in this work, a decimal decomposition provides a simple and transparent way to navigate the hierarchy, with an intuitive correspondence between digits and successive refinement levels. We retained this choice throughout the study because adopting a different base would require a systematic analysis of the trade-off between reducing the number of algebraic comparisons at each level and increasing the number of cumulative probability vectors that must be constructed, stored, and updated. Since such an optimization study lies beyond the scope of the present work, and no clear performance gain could be anticipated without it, we chose to keep the base-10 implementation as a well-defined and effective reference. Nevertheless, to assess the sensitivity of the algorithm to this choice, we performed additional tests on 100 independent magnetization reversal simulations using a base-2 hierarchical construction with the same physical parameters already used in our studies ($L = 100$, $H = 1.5$, $\sigma^2$ ranging from $0.1$ to $0.5$ and temperature $T$ ranging from $0.2$ to $2.7$). We found that the resulting wall-clock times differed from the base-10 implementation by less than a factor of two, with a maximum variation of approximately $\pm35\%$, depending on parameters. This confirms that the overall performance of the algorithm is only weakly dependent on the specific base chosen, and that the base-10 implementation provides a representative benchmark.

Given a system of $N$ spins, we let $n = \text{int} \left[\log_{10} \left(N -1\right)\right]$ be the largest integer such that $10^n \leq N$, and build an array of length $\lfloor \frac{N - 1}{10^n}\rfloor + 1$ in which the i-th element contains the sum of the transition probabilities of the spins in the lattice positions between $\left( i-1 \right) \times 10^n$ and $i \times 10^n -1$. The same procedure is repeated for the power $10^{n-1}$ by constructing an array of length $\lfloor \frac{N - 1}{10^{n-1}}\rfloor + 1$ containing the sum of $10^{n-1}$ transition probabilities in each of its elements. This recursive method goes on until we create the array corresponding to the power $10^n$ containing every spin's transition probability. After creating all these arrays, we extract a random real number in the $ \left[ 0,p_{tot} \right] $ interval and we call it \textit{random} (as in the actual code). We start by comparing \textit{random} to the value contained in the probability counter of power $n$ and position $i = 0$: if \textit{random} is the lower of the two, we can say that the selected spin we need to identify is among the first $10^n$ spins. Otherwise, we should compare \textit{random} to the sum of the two counters identified by $i = 0$ and $i = 1$: if \textit{random} is lower, then the selected spin belongs to the segment containing the second $10^n$ spins (from $1 \times 10^n$ and $2 \times 10^n - 1$), otherwise we should go on and compare \textit{random} to the sum of the elements in the first three counters ($i$ up to 2) and so on. The highest value of the counter $i$ included in the first sum larger than \textit{random} will be the first of the $n$ digits we need to identify the spin to flip. This counter \textit{i} identifies the interval of spins in which the selected spin is located. Once we have found the right interval using this first counter, we repeat the same procedure into the selected interval, identifying 10 intervals into that, giving the sum of $10^{n - 1}$ spins' transition probabilities each. Then, we go on with the same reasoning to find the second digit and the following ones, iterating this process until we find the unit digit. A schematic representation of this procedure of building and using of the hierarchical cumulative counter vectors is shown in Fig. \ref{fig:Flowchart}. Hence, with a few, simple, algebraic comparisons, we are able to randomly select a spin in a way proportional to the transition probability of each spin in the lattice. At most, this method would require making $9n$ comparisons for each spin selection, a significant improvement over the Metropolis algorithm, which nominally incurs $\mathcal{O}(1)$ computational cost per attempted move, but, at low temperatures, the acceptance probability in Metropolis becomes exceedingly small and the system experiences long "waiting times" in configuration space, leading to a substantially increased wall-clock time per accepted event. Moreover, once we flip the selected spin, we just need to update the energy and the transition probability of the flipped spin and of its four nearest neighbors only, then proceed to update the cumulative counters by simply adding to them the difference between the transition probability of these 5 spins (the flipped one and its 4 nearest neighbors) after and before the actual spin flip.

\section{Validation and Performance Comparison with Metropolis}
\label{sec: Validation and Performance Comparison with Metropolis}

\subsection{Validation}
\label{subsec: Validation}

\begin{figure}[t]
  \centering
  \includegraphics[width=0.5\textwidth]{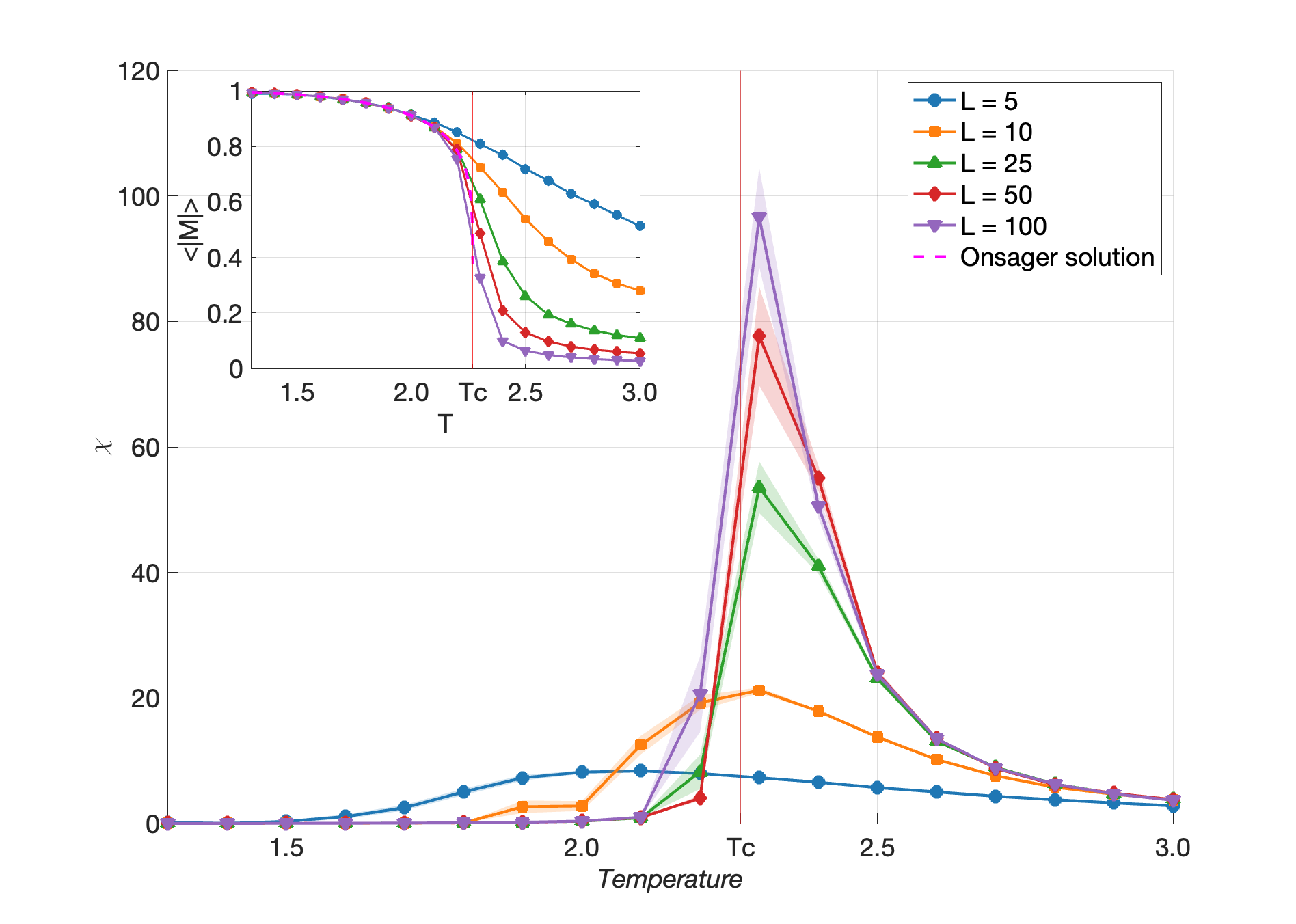}
  \caption{Inset: Average magnetization per site $<|M|>$ versus temperature $T$ for different lattice sizes $L$ for the 2D Ising Model with no RF applied.\\
  Main figure: Susceptibility $\chi$ versus temperature $T$ for different lattice sizes $L$ for the 2D Ising Model with no RF applied.\\
  Circles in the figure represent the average of the simulation results. A solid line connects the simulation values for visualization purposes, while the areas around it show the standard error of the mean.}
  \label{fig:Chi_M_Vs_T_0_ErrBars}
\end{figure}

The first analysis to test the correct functioning of this algorithm was carried out by setting the RF to zero (i.e., the mean and the variance of the random component of the field are set to zero) and simulating a pure Ising System to see if the outcomes complied with Onsager's analytical results \cite{Onsager1944}. All algorithms in this work were implemented in the \textit{C++} programming language. This choice was motivated by its efficiency and low-level control over memory management, which are crucial for large-scale Monte Carlo simulations. All simulations were performed on hardware with a Intel Core i7-4700MQ CPU @2.40GHz. In this simulation, 10 different lattice configurations, initialized at random, have been studied under the effect of 10 different RF configurations, for a total of 100 simulated systems for each lattice size and temperature values. A lattice size $L = 100$ was used, with a simulation time = 100000, which is much larger than the typical requirement for a $100 \times 100$ spin system to reach equilibrium in the studied temperature range. The first half of the simulation time was used for thermalization, while the actual measurements were taken in the second half. To account for temporal correlations in the Markov chain, a blocking (binning) analysis was applied to the time series before computing the realization average. The disorder-averaged observable was then obtained by averaging over all realizations. Error bars correspond to the standard error of the mean (SEM) across realizations. The results obtained for the average magnetization as a function of the temperature are reported in the inset of Fig. \ref{fig:Chi_M_Vs_T_0_ErrBars}, showing excellent agreement with the Metropolis-based numerical data from Fig. 2 of Ref. \cite{Ibarra-Garca-Padilla2016}. This plot clearly shows that the magnetization approaches the shape predicted by Onsager's solution at the thermodynamic limit, given by: 
\begin{equation}\label{eq.Onsager}
    \frac{M}{N} = \begin{cases}
    0 & \text{for } T > T_c, \\[10pt]
    \pm \left[1 - \sinh^{-4}(2\beta J)\right]^{1/8} & \text{for } T < T_c,
    \end{cases}
\end{equation}
shown in the inset of Fig. \ref{fig:Chi_M_Vs_T_0_ErrBars} by the magenta dotted line. 

\begin{figure}[]
  \centering
  \includegraphics[width=0.5\textwidth]{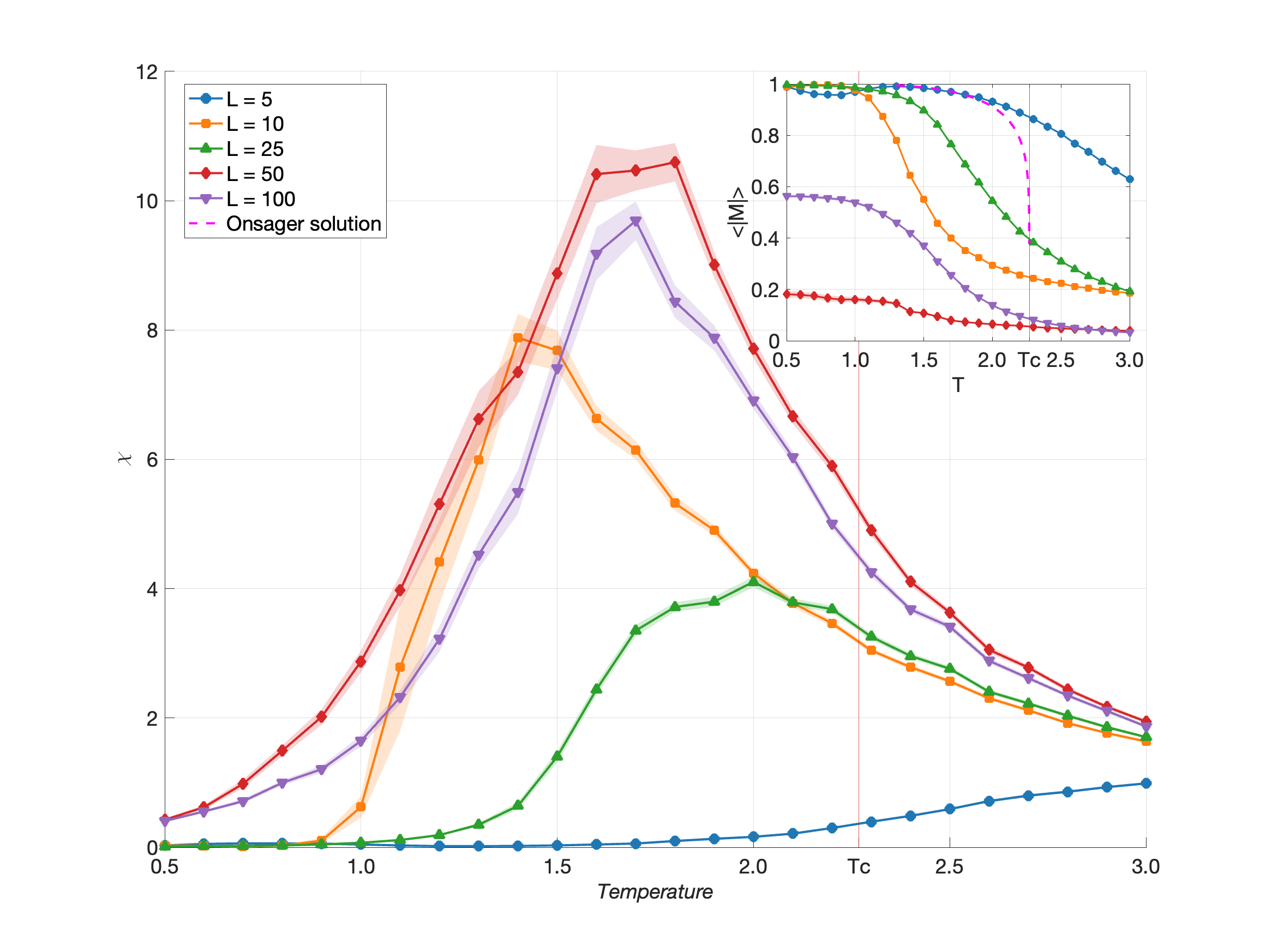}
  \caption{Inset: Average magnetization per site $<|M|>$ versus temperature $T$ for different lattice sizes $L$ for the 2D RFIM with $\sigma = 1.0$.\\
  Main figure: Susceptibility $\chi$ versus temperature $T$ for different lattice sizes $L$ for the 2D RFIM with $H = 0$ and $\sigma^2 = 1.0$.\\
  Circles in the figure represent the average of the simulation results. A solid line connects the simulation values for visualization purposes, while the areas around it show the standard error of the mean.}
  \label{fig:Chi_M_Vs_T_1.0_ErrBars}
\end{figure}
This first analysis shows that the algorithm correctly reproduces the final state reached by the Ising system at different temperatures. Another fundamental property the algorithm must have is the ability to find the critical temperature value for the 2D Ising Model. This assessment is usually performed by evaluating the susceptibility values as the temperature changes, seeking for a clearly defined peak. The susceptibility $\chi$ is calculated as:
\begin{equation}\label{eq.chi}
    \chi = \frac{\beta}{N}\left(<M^2> - <M>^2\right)
\end{equation}
The behavior of the susceptibility is represented in Fig. \ref{fig:Chi_M_Vs_T_0_ErrBars}, where we can clearly see the peak in susceptibility indicating the critical temperature value. Due to finite size effects, in finite systems the critical values will not be the same as in the thermodynamic limit, but we can find excellent agreement with the results of other computational studies, as for Fig. 1 of \cite{AbdulRazak2014}. Therefore, we can confirm that this code works according to plan if we study an Ising system with no RF applied. 

However, this algorithm was developed to work correctly even in the presence of a RF applied. If we consider a RF with a normal distribution centered in 0 and a small variance value $\sigma^2$, e.g. using $\sigma^2 = 0.01$, the results regarding the average magnetization and susceptibility are essentially the same of the pure Ising model with no RF applied. This is due to the fact that this value of the RF is not strong enough to lower the critical temperature value or to create clustered equilibrium configurations for systems of lateral size $L = 100$ like the ones considered here.

Instead, by simulating 100 different lattices, initialized at random, under the effect of a RF with zero mean and variance $\sigma^2 = 1.0$, we observe that, even after long simulation times, no full magnetization is reached. Moreover, the susceptibility curve develops a broader and less pronounced peak, as shown in Fig. \ref{fig:Chi_M_Vs_T_1.0_ErrBars}, and a clear drift of the curve maximum to lower temperatures. This indicates a decrease in the critical temperature value when the disorder in the system, i.e. the standard deviation $\sigma$ of the applied RF, increases, as already reported by \cite{Metra2021}. 

\begin{figure}[t]
  \centering
  \includegraphics[width=0.5\textwidth]{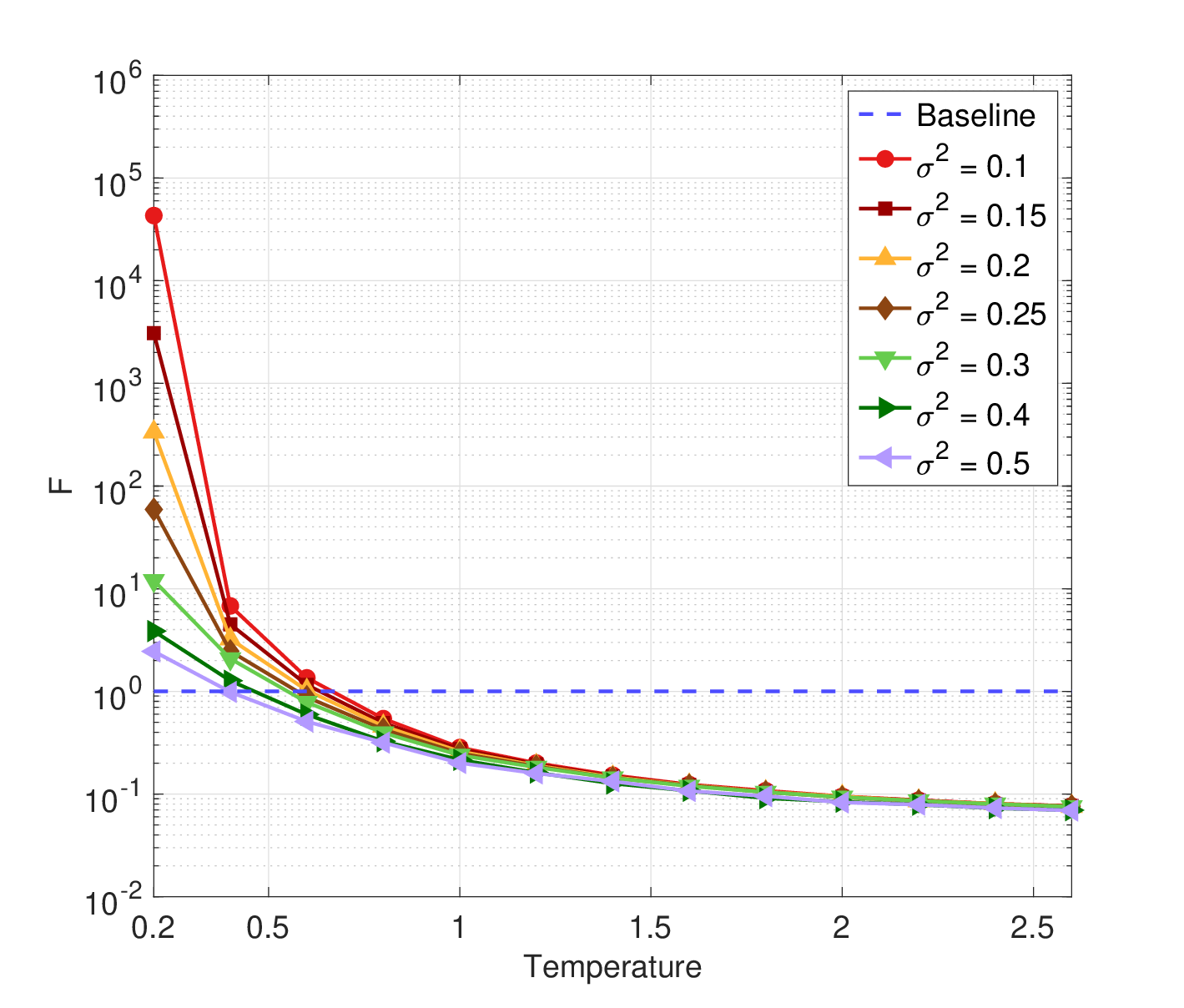}
  \caption{Speedup Factor $F$ in the simulation of the magnetization reversal of a RFIM with lattice size $L = 100$ and an applied Gaussian RF with $H = 1.5$. Different colors show the performance when different variance values $\sigma^2$ of the applied field are used. \\
  Circles in the figure show the average of the simulation results, while a solid line connects the simulation values for visualization purposes.}
  \label{fig:Performance_100_15}
\end{figure}

\begin{figure}[]
  \centering
  \includegraphics[width=0.5\textwidth]{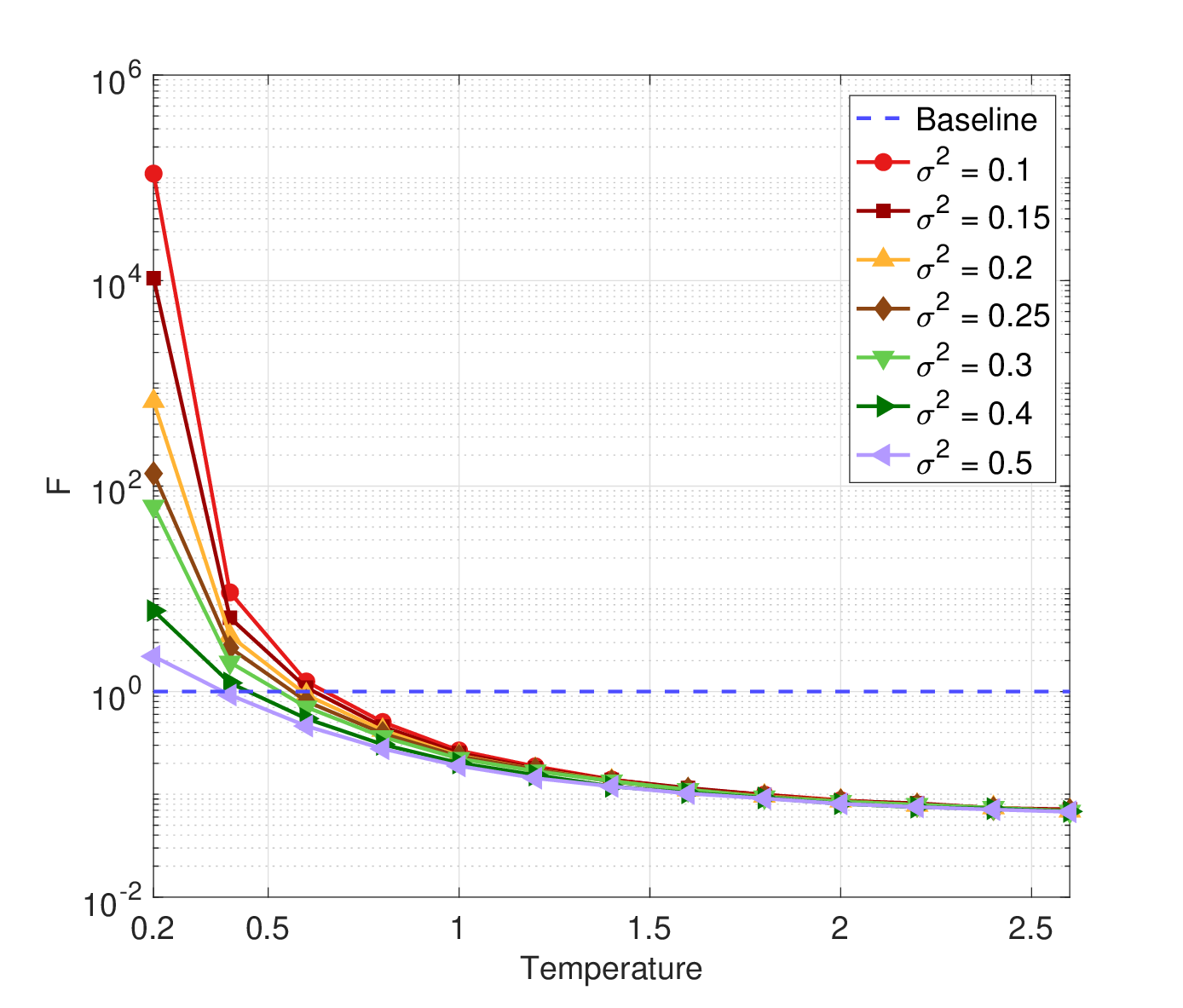}
  \caption{Speedup Factor $F$ in the simulation of the magnetization reversal of a RFIM with lattice size $L = 60$ and an applied Gaussian RF with $H = 1.5$. Different colors show the performance when different variance values $\sigma^2$ of the applied field are used. \\
  Circles in the figure show the average of the simulation results, while a solid line connects the simulation values for visualization purposes.}
  \label{fig:Performance_60_15}
\end{figure}

\subsection{Performance comparison}
\label{subsec: Performance comparison}

Before discussing the choice of the simulation protocol, it is important to note that a direct performance comparison with the standard N-fold way (BKL) algorithm is not meaningful in the present context. As discussed above, the presence of a site-dependent random field prevents an efficient class-based organization of spins, so that a naïve BKL implementation would require
$\mathcal{O}(N)$ updates of the transition probabilities after each spin flip, effectively removing the algorithm’s computational advantage. For this reason, the performance analysis presented here focuses on a comparison with the Metropolis algorithm, for which a well-defined and efficient implementation in the RFIM is available.

To choose the best simulation setup to test the performance of this algorithm, we first need to notice some peculiarities of the code. Using an a-priori approach and the Glauber dynamics, this algorithm, like the N-Fold Way, allows us to utilize either the \textit{MCSS} (Monte Carlo Steps per Spin, as in the Metropolis algorithm) or the \textit{physical time} as a reference for the time evolution in the simulation. In the Metropolis algorithm, the MCSS resembles the physical-time condition, since the probability of remaining in the same state $P(H \to H) \neq 0$: at each simulation step does not necessarily correspond the creation of a new state and for moves (i.e. spin flips) that bring the system far from equilibrium, more simulation steps are required, since more moves will be discarded. Instead, the MCSS in the N-Fold Way algorithm follows the succession of states in the Markov chain, without considering the possibility that the time for the transition between two states could vary. Indeed, since the acceptance probability is always equal to 1, there will always be one simulation step between a state and the next. But this does not seem to be physically very accurate. For this reason, we consider the physical time as the time reference in our studies, as well as in continuous-time rejection-free algorithms. This means that, to compare the performance of this algorithm to that of the Metropolis, we have to use a method that is not limited by a maximum evolution time, otherwise we would need to find a relation between MCSS and physical time in the temperature range studied, and this goes beyond the interest of this analysis. Hence, to avoid this problem, we set the initial lattice to completely negative magnetization (all spins with negative orientation) and initialize 10 different RF configurations using a Gaussian distribution with the same $H$ and $\sigma^2$ values. Then, for each RF realization, we perform 10 different repetitions of the simulation, in order to have 100 different simulations for each combination of the lattice size $L$ and the parameters of the RF's distribution $H$ and $\sigma$, and this is repeated at each studied temperature value. The aim is to observe the system reverse its magnetization under sufficiently strong RFs. So, we stop the simulation when the system exhibits $M \geq 0$ value on at least 3 consecutive spin flip events, in order to ensure a simulation times' comparison as objective as possible. We emphasize that this benchmark is not intended to provide an exhaustive performance analysis over all possible parameter combinations or initial conditions; rather, it is designed to highlight the regime in which the standard Metropolis approach becomes inefficient and to illustrate the computational advantages of the proposed algorithm under low-temperature, low-disorder conditions.

\begin{figure}[]
  \centering
  \includegraphics[width=0.5\textwidth]{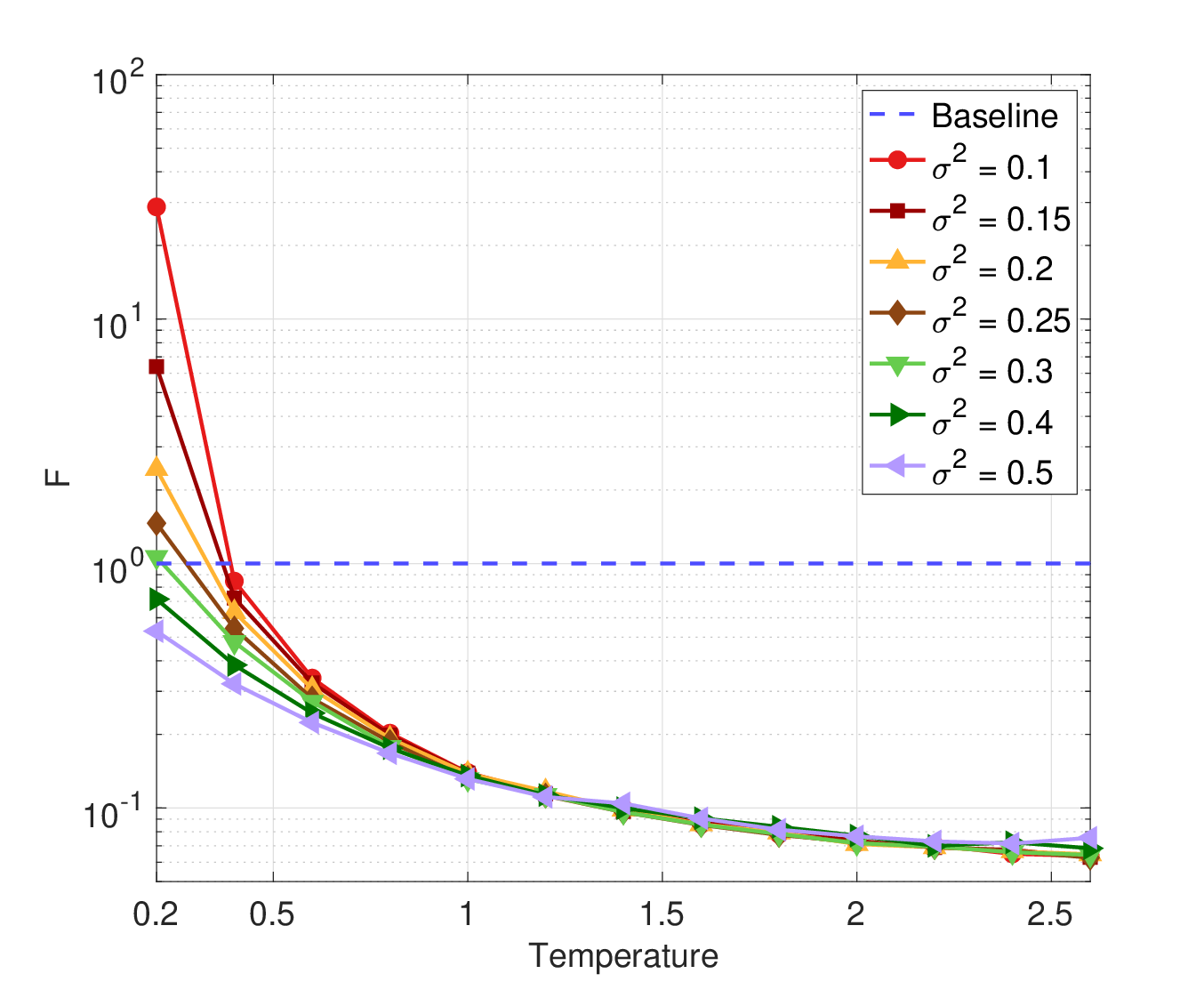}
  \caption{Speedup Factor $F$ in the simulation of the magnetization reversal of a RFIM with lattice size $L = 100$ and an applied Gaussian RF with $H = 2.0$. Different colors show the performance when different variance values $\sigma^2$ of the applied field are used. \\
  Circles in the figure show the average of the simulation results, while a solid line connects the simulation values for visualization purposes.}
  \label{fig:Performance_100_20}
\end{figure}

\begin{figure}[]
  \centering
  \includegraphics[width=0.5\textwidth]{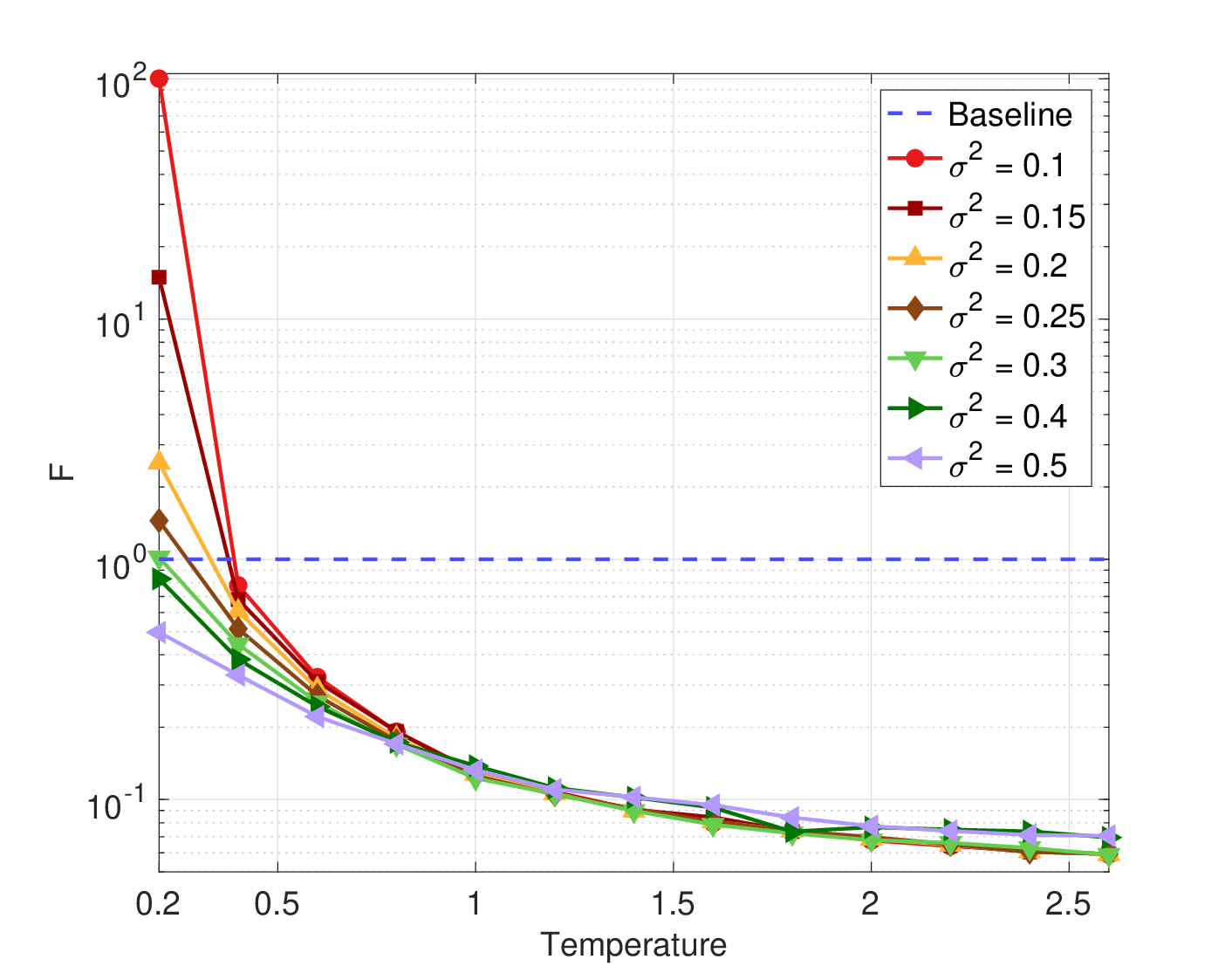}
  \caption{Speedup Factor $F$ in the simulation of the magnetization reversal of a RFIM with lattice size $L = 60$ and an applied Gaussian RF with $H = 2.0$. Different colors show the performance when different variance values $\sigma^2$ of the applied field are used. \\
  Circles in the figure show the average of the simulation results, while a solid line connects the simulation values for visualization purposes.}
  \label{fig:Performance_60_20}
\end{figure}

Fig. \ref{fig:Performance_100_15} shows the Speedup Factor $F$ defined as:
\begin{equation}\label{Eq:Speedup Factor}
    F = \frac{T_{Metropolis}}{T_{New}}
\end{equation}
where $T_{Metropolis}$ is the CPU wall-clock time needed to simulate the magnetization reversal of the RFIM using the Metropolis algorithm on a single core, while $T_{New}$ is the time required by the algorithm introduced in the present work using the same hardware. A value greater than 1 means that the present algorithm outperforms the Metropolis algorithm. For all $\sigma^2$ values, the Metropolis approach becomes slower at sufficiently low temperatures. Moreover, the smaller the disorder (i.e., the $\sigma^2$ value of the RF distribution), the higher is the gain in terms of computational time saved by using the present algorithm. This can be understood by the fact that the Metropolis algorithm at low temperature rejects many moves at each step, but a higher disorder means that it will be easier for it to randomly find lattice points with high enough local values of the RF. This increases the transition probability for the spins in those lattice positions and reduces the time needed to the Metropolis algorithm to exit those unfavorable ordered configurations.  

Applying a RF with the same probability distribution's parameters $H$ and $\sigma^2$ to a system of size $L = 60$, we see in Fig. \ref{fig:Performance_60_15} a similar trend in the data collected, with a general increase in the values of the speedup factor at each studied temperature. This indicates better performance of the studied algorithm with respect to the Metropolis one and a different performance scaling with the linear lattice dimension $L$ between the two approaches, and this will be explored further later in this Section.

The same analysis has been repeated using a RF distribution with $H = 2.0$ on lattices of the same previous dimensions $L = 60$ and $L = 100$. In Fig. \ref{fig:Performance_100_20} and \ref{fig:Performance_60_20} we can observe that there is not a great change in the performance trends of the two algorithms with temperature. However, with the increase of the $H$ value of the RF distribution, we can notice a reduction in the relative performance of the present algorithm. This is due to the fact that we are now applying a field which is, on average, greater than the previous one, and this helps the Metropolis algorithm find spins with high acceptance probability. In fact, with the same values of the variance of the RF, a higher mean value means that it will be more likely to reach larger local RF values on some spins of the lattice, which will show a much higher probability to be flipped in the Metropolis approach. Instead, the present algorithm flips a spin at each move, and the spin's finding time (therefore the computational time) is not related to the energy of the spins in the lattice. This means that a higher $H$ value reduces the gain in the computational time saved by using the present algorithm with respect to the Metropolis one.

\begin{figure}[]
  \centering
  \includegraphics[width=0.55\textwidth]{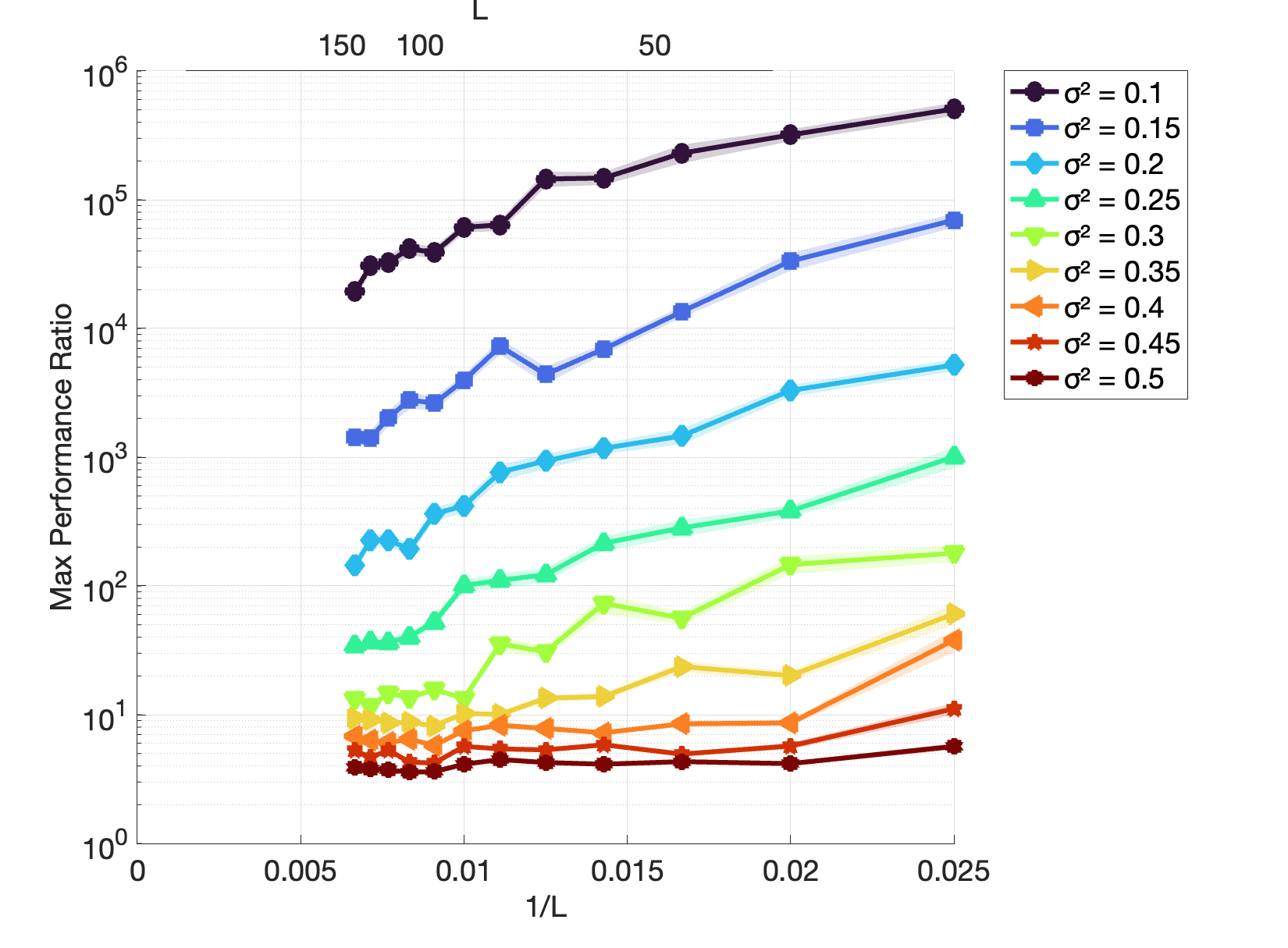}
  \caption{Max Speedup factor $F_{max}$ with a Gaussian RF applied with $H = 1.5$, at different lattice sizes. Different colors correspond to different variance values $\sigma^2$ of the applied RF. \\
  Circles in the figure represent the average of the simulation results. A solid line connects the simulation values for visualization purposes, while the small areas around it show the standard error of the mean.}
  \label{fig:RatioVsL_quart_1.5_L}
\end{figure}

\begin{figure}[t]
  \centering
  \includegraphics[width=0.5\textwidth]{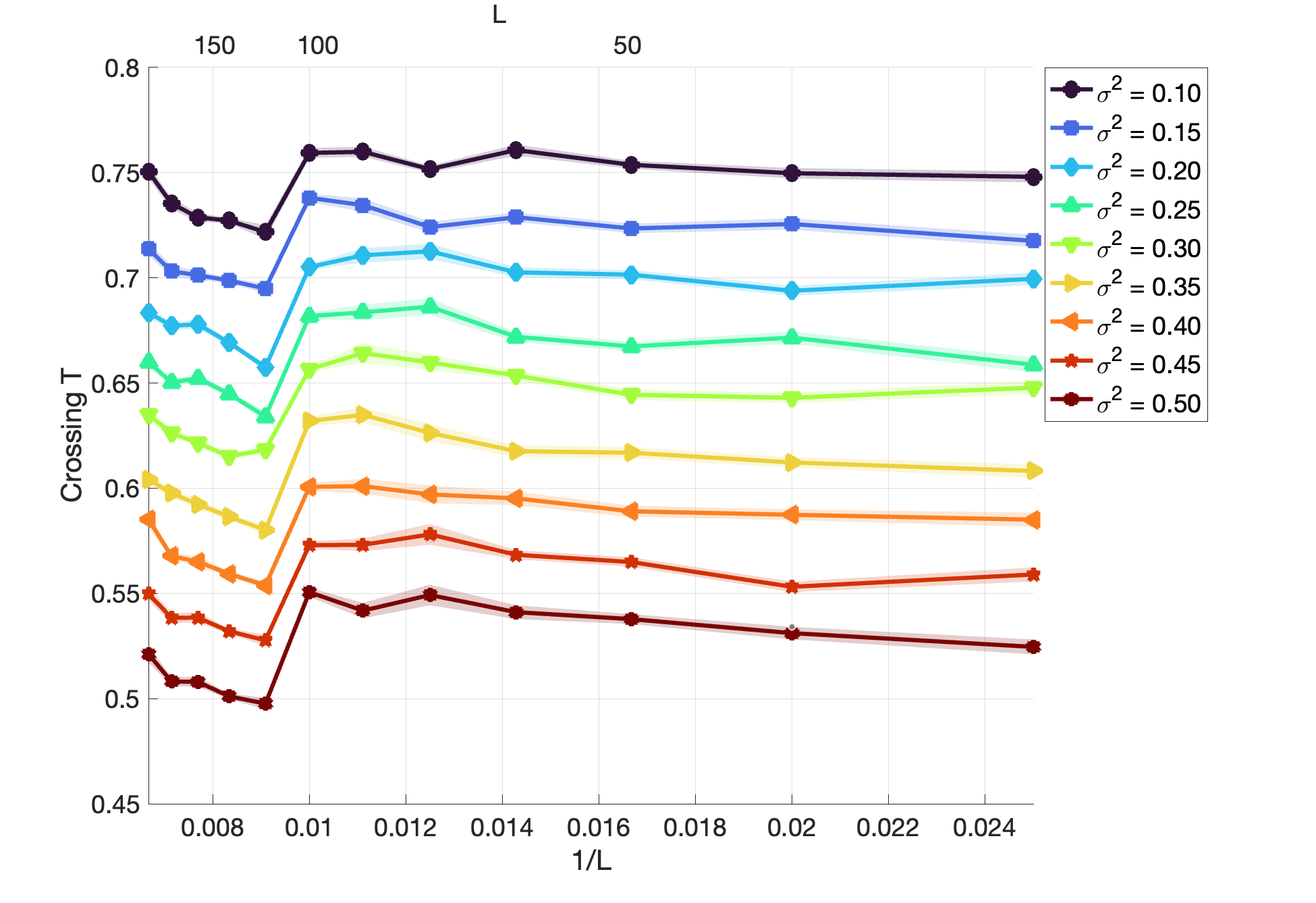}
  \caption{Temperature value at which the present algorithm shows the same performance of the Metropolis one when a Gaussian RF with $H = 1.5$ is applied, at different lattice sizes. Different colors correspond to different variance values $\sigma^2$ of the applied RF. \\
  Circles in the figure represent the average of the simulation results. A solid line connects the simulation values for visualization purposes, while the small areas around it show the standard error of the mean.}
  \label{fig:TcrossVsL_quart_1.5_L}
\end{figure}

We do not show performance studies at lower values of temperature (below 0.2) or $H$ (below 1.5) because Metropolis simulations with these parameters are too demanding for our actual purposes, in terms of computational cost and time required. In fact, a larger applied field facilitates magnetization reversal, while at smaller $H$ the reversal dynamics becomes extremely slow, particularly for the Metropolis algorithm, whose performance strongly depends on the local energy changes of individual spin flips. As a result, simulations at low $H$ and low $T$ are too computationally demanding when using Metropolis dynamics. The choice of the field values used in this benchmark is therefore dictated by practical constraints on wall-clock time. We attempted simulations at lower field values (e.g., $H = 1$), but already at $T = 0.3$ the wall-clock time required by the Metropolis algorithm became prohibitive, due to both the slower magnetization reversal dynamics and the sharply reduced acceptance rate of Metropolis updates. Alternative benchmarks could provide complementary information, the present benchmark is designed to emphasize the regime where standard Metropolis approach becomes impractical, and where the advantages of the proposed algorithm are most pronounced.

To further analyze the convenience of the present algorithm compared to the Metropolis one, in Fig. \ref{fig:RatioVsL_quart_1.5_L} are shown the maximum values of the speedup factor $F_{max}$ at each studied lattice size $L$. The maximum $F$ value is always reached at the lowest studied temperature ($T = 0.2$), where the Metropolis approach slows down significantly. As we can observe, the algorithm introduced above shows clear performance gains at every lattice size and RF variance. As in previous analyses, here too we observe the same increase in performance at lower RF variance, while we cannot find a clear trend as a function of the lattice dimension $L$.  

\begin{figure}[]
  \centering
  \includegraphics[width=0.5\textwidth]{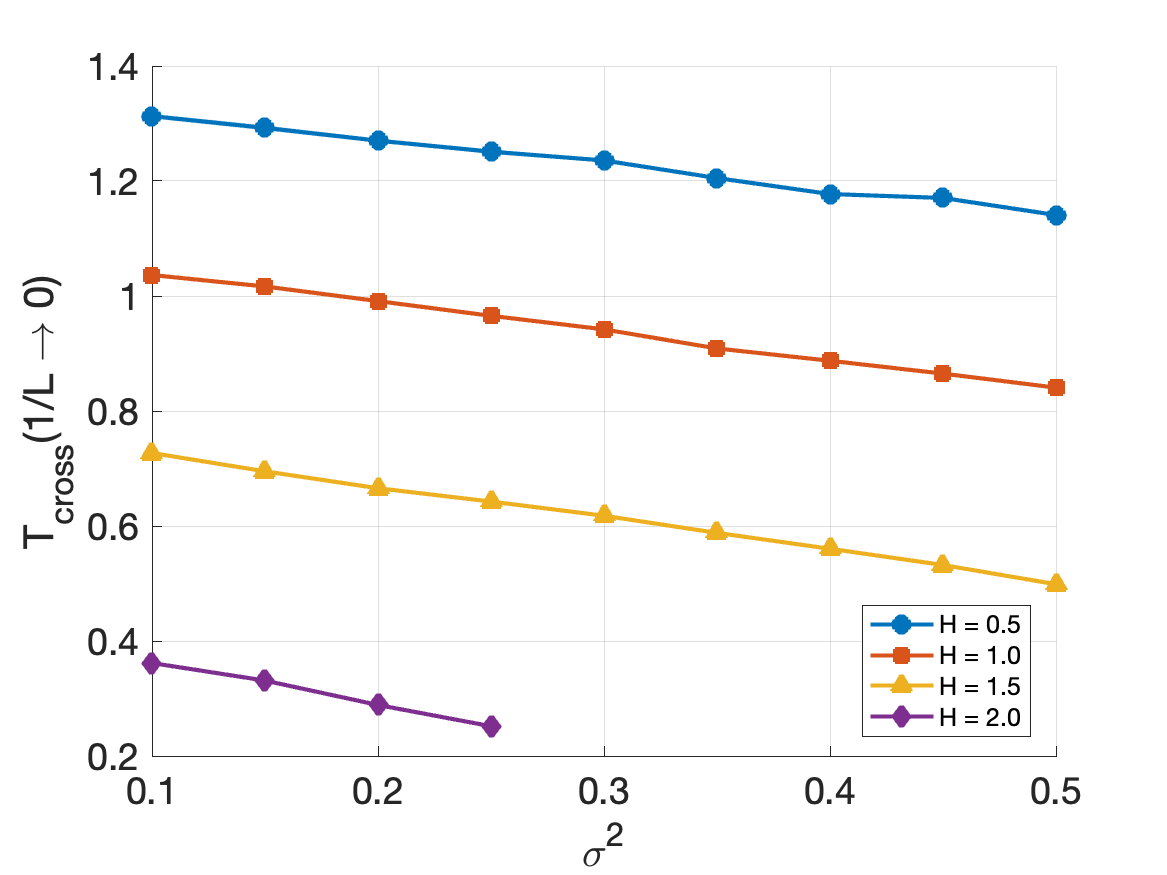}
  \caption{Extrapolated temperature values at the thermodynamic limit ($1/L \to 0$) at which the present algorithm shows the same performance of the Metropolis in the simulation of a RFIM with Gaussian RF applied.\\
  Circles in the figure represent the average of the simulation results. A solid line connects the simulation values for visualization purposes, while the small areas around it show the standard error of the mean.}
  \label{fig:TcrossVsB_tot_final}
\end{figure}

Then, in Fig. \ref{fig:TcrossVsL_quart_1.5_L} we find the interpolated temperature values at which the two approaches show the same performance, as a function of the inverted linear lattice dimension 1/$L$. These values are obtained through interpolation using results like the ones of Fig. \ref{fig:Performance_100_15} and \ref{fig:Performance_60_15}. Also in this case, the performance dependence on the lattice size $L$ and the RF variance $\sigma^2$ is the same. In fact, a higher crossing temperature at lower RF variance is somehow correlated to the higher value of the speedup factor at $T = 0.2$ found above. If the performance curve of Fig. \ref{fig:Performance_100_15} crosses the $10^0$ value at higher temperature, we expect the speedup factor at $T = 0.2$ to be higher with respect to the case of lower crossing temperature value, because of the typical shape of these performance curves. To assess the possible bias associated with the interpolation procedure used to determine the crossing temperatures, we explicitly tested the quality of the interpolant. In particular, we performed the analysis using two different interpolations of the averaged data: a linear interpolant and a spline interpolant. We found that the resulting crossing temperatures obtained from the two methods are nearly indistinguishable within the statistical uncertainties, indicating that the interpolation bias is negligible at the resolution of the present data and not responsible for the observed fluctuations. Based on this comparison, we retained the linear interpolation of the averaged values in Fig. \ref{fig:TcrossVsL_quart_1.5_L} for simplicity. 

A closer inspection of the crossing temperatures corresponding to $L = 100$ and $L = 110$ reveals something that seems a small but systematic shift in the estimated crossing temperature values for all considered disorder strengths. This feature can be traced back to an implementation-specific aspect of the proposed algorithm discussed in Sec. \ref{sec: The algorithm}, namely the use of a base-10 hierarchical reference in the construction of the cumulative counters. For $L = 100$, the lattice contains $N = 10000$
spins (indexed from 0 to 9999), which can be uniquely addressed using four decimal digits in the hierarchical selection procedure. Increasing the system size to $L = 110$ ($N = 12100$) requires an additional digit in the base-10 representation, leading to a change in the depth of the hierarchical structure and, consequently, in the algorithm’s effective performance. Since this implementation detail does not affect the Metropolis algorithm, whose computational cost scales smoothly with system size, the temperature at which the two methods exhibit comparable performance is correspondingly shifted.

\begin{figure}[]
  \centering
  \includegraphics[width=0.5\textwidth]{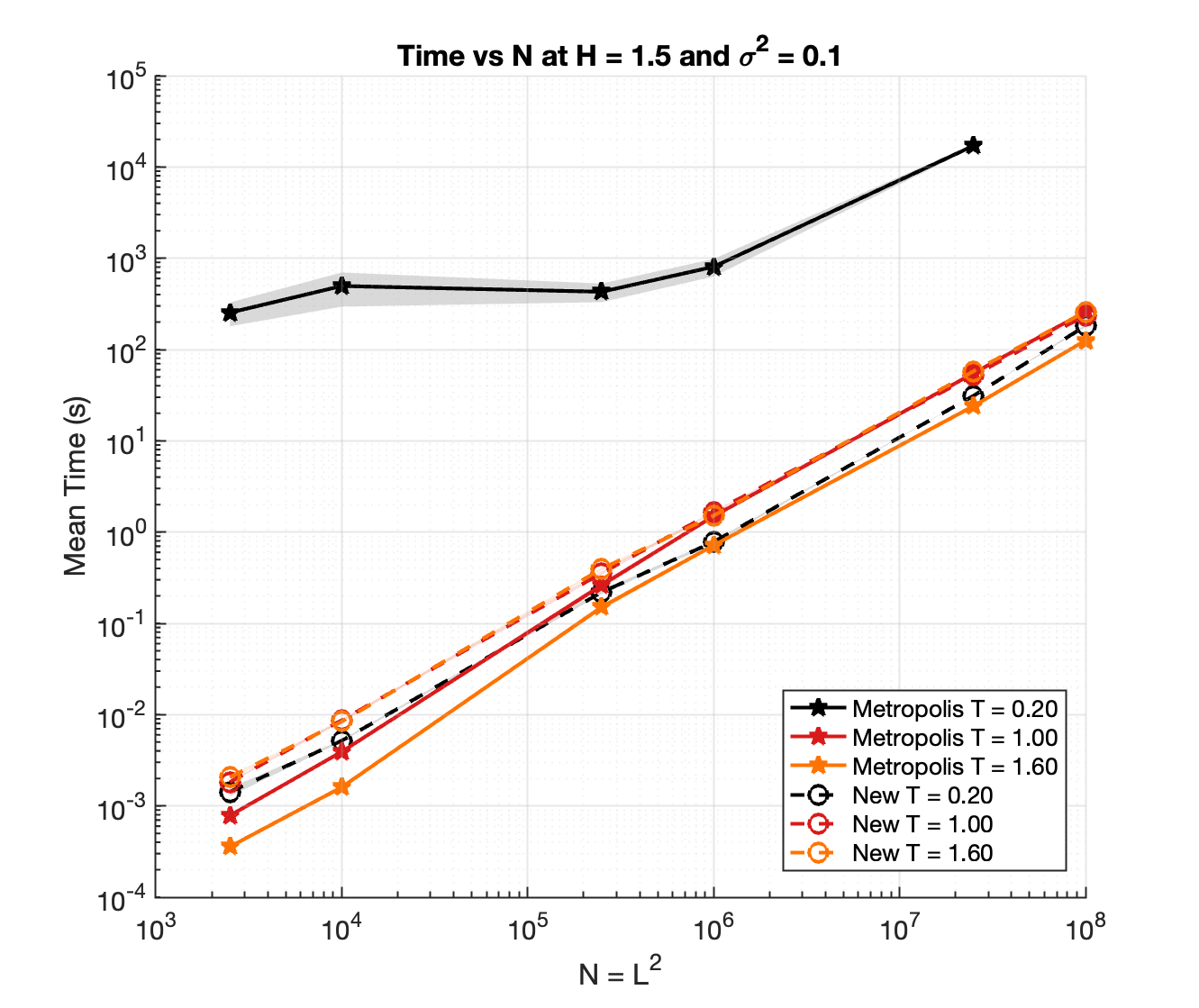}
  \caption{Wall-clock time needed by the two algorithms analyzed in this study to simulate the magnetization reversal of a RFIM at different temperatures and with different lattice dimensions $N$, with fixed values of $H = 1.5$ and $\sigma^2 = 0.1$. \\
  Circles in the figure represent the average of the simulation results. A solid line connects the simulation values for visualization purposes, while the small areas around it show the standard error of the mean.}
  \label{fig:TimeVsL_0.1_1.5_tot}
\end{figure}

By performing the same linear fit as the one done in Fig. \ref{fig:RatioVsL_quart_1.5_L} for the $F$ values, this time for the crossing temperatures of Fig. \ref{fig:TcrossVsL_quart_1.5_L}, we find the extrapolated $T_{cross}$ at different RF variance and H values shown in Fig. \ref{fig:TcrossVsB_tot_final}. These values represent the extrapolated temperature value at the thermodynamic limit ($1/L \to 0$) at which the two algorithms show the same performance in terms of computational time needed. We can clearly notice that at $H = 2.0$ we find fewer values with respect to the other cases. This is related to the fact that a higher H value of the applied RF gives, on average, a higher energy to the spins in the lattice, helping the Metropolis algorithm to improve its performance. Therefore, in the case of high H values and high RF variance (i.e. high disorder), there is no crossing temperature because the Metropolis algorithm outperforms the present algorithm at any studied temperature (i.e. any $T \geq 0.2$). In the same way, the crossing temperature decreases when we increase $H$ or $\sigma^2$, as we can see in the 4 curves in Fig. \ref{fig:TcrossVsB_tot_final} corresponding to different $H$ values. The numerical values of variance $\sigma^2$ and $H$ for which we no longer see this crossing temperature value may vary with the size of the system $N$, which was fixed to $10^4$ ($L = 100$) in this initial analysis.

Finally, we analyzed the computational time behavior of this algorithm when increasing the lattice size $N = L^2$ in the magnetization reversal simulations used for the results above, compared to the Metropolis one. In Fig. \ref{fig:TimeVsL_0.1_1.5_tot}, we find the wall-clock time required by the two algorithms to simulate the magnetization reversal of a RFIM with fixed $H = 1.5$ and $\sigma^2 = 0.1$ and variable lattice dimension. As we can notice, the two algorithms show a very similar $\mathcal{O}(N)$ scaling of the wall-clock time on $N$, and this happens for each studied value of $H$ and $\sigma^2$. This is due to the nature of the physical process studied, which needs at least $N/2$ spin flips to complete the magnetization reversal. So, the $\mathcal{O}(N)$ scaling is the best result we can expect in this kind of simulations. However, things change approaching lower temperatures. In fact, at low temperature ($T = 0.2$) the Metropolis algorithm shows a huge slowdown, due to its rejection rate based on energetic evaluations. This slowing down effect is obviously enhanced at lower $H$ and $\sigma^2$ values, because of the lower average energetic contribution given by the RF to the spins, and scales as $\mathcal{O}(\frac{1}{p_{acc}})$, where $p_{acc}$ is the Metropolis acceptance probability. This acceptance rate does not depend on the lattice size $N$, so the slowing down due to low $T$ is almost equal for any $N$ value studied here. At small $N$, this slowdown dominates the computational time, whereas at larger $N$ the $\mathcal{O}(N)$ scaling becomes the dominant contribution. Instead, with larger lattices, we have a visible wall-clock time increase due to the larger lattice dimension $N$. 

This simulation setup therefore provides a clear example of the relevance of the proposed algorithm when dealing with low temperature, low disorder RFIM. Working with a RFIM, the BKL algorithm cannot exploit a reduced number of classes, leading to a substantial loss of its characteristic speedup, while the Metropolis algorithm in this regime suffers from a severe critical slowdown due to extremely small acceptance rates. The proposed algorithm avoids both limitations, maintaining an efficient performance precisely in the parameter region where standard approaches become impractical.

\section{Conclusions}
\label{sec: Conclusions}

We have developed an alternative algorithm for the simulation of the 2D RFIM, which significantly improves performance over the standard Metropolis algorithm in the low-temperature regime. Moreover, because this approach is based on Glauber dynamics, it can properly describe the dynamics of the system, both in equilibrium and out of equilibrium. Starting from the approach of the BKL algorithm, but considering the presence of the RF, which radically changes the description of the system and removes the possibility of grouping all spins into a small number of classes, this algorithm correctly captures both the magnetization and susceptibility behavior of the RFIM, showing also a clear lowering of the critical temperature value when increasing the disorder (i.e. the variance of the RF distribution) in the system, as expected from previous works (see \cite{Metra2021}). The performance analysis performed on RFIMs with Gaussian distributions of the applied RF shows an average gain of two orders of magnitude (or more) in the computational time with respect to the Metropolis algorithm at low temperature, even in the presence of significant disorder (i.e. RF with variance $\sigma^2 = 0.5$). 

Future applications of this algorithm range from the study of the RFIM at temperatures lower than those studied so far, also with different RF distributions applied, to the study of the dynamic behavior of the RFIM in the low-temperature regime, thanks to the fact that this method is based on the Glauber dynamics.

\subsection*{Author contributions} 
\begin{itemize}
    \item Luca Cattaneo: Conceptualization, Algorithm development, Validation, Performance study, Writing – original draft, Writing – review $\&$ editing.
    \item Federico Ettori: Conceptualization, Writing – review $\&$ editing.
    \item Giovanni Cerri: Conceptualization, Algorithm development, Validation, Writing – review $\&$ editing.
    \item Paolo Biscari: Conceptualization, Writing – review $\&$ editing.
    \item Ezio Puppin: Conceptualization, Writing – review $\&$ editing.
\end{itemize}

\subsection*{Data availability} The data generated and analyzed during this study are available upon reasonable request from the corresponding author.

\subsection*{Code availability} The code developed during this study can be obtained upon reasonable request from the corresponding author.

\subsection*{Conflict of interest} The authors declare that they have no financial or personal relationships that could be perceived as influencing the research reported in this manuscript.

\begin{appendices}

%%=============================================%%
%% For submissions to Nature Portfolio Journals %%
%% please use the heading ``Extended Data''.   %%
%%=============================================%%

%%=============================================================%%
%% Sample for another appendix section			       %%
%%=============================================================%%

%% \section{Example of another appendix section}\label{secA2}%
%% Appendices may be used for helpful, supporting or essential material that would otherwise 
%% clutter, break up or be distracting to the text. Appendices can consist of sections, figures, 
%% tables and equations etc.

\end{appendices}
%%===========================================================================================%%
%% If you are submitting to one of the Nature Portfolio journals, using the eJP submission   %%
%% system, please include the references within the manuscript file itself. You may do this  %%
%% by copying the reference list from your .bbl file, paste it into the main manuscript .tex %%
%% file, and delete the associated \verb+\bibliography+ commands.                            %%
%%===========================================================================================%%

\bibliography{sn-bibliography}% common bib file
%% if required, the content of .bbl file can be included here once bbl is generated
%%\input sn-article.bbl

\end{document}